# Single crystal growth and characterization of $Ba_2ScNbO_6$
## - a novel substrate for $BaSnO_3$ films

C. Guguschev [a,*], D. Klimm [a], M. Brützam[a], T. M. Gesing [b,c],
M. Gogolin [b], H. Paik [d,e], A. Dittmar [a], V. J. Fratello [f], D.G. Schlom [d,g]

[a] Leibniz-Institut für Kristallzüchtung, Max-Born-Str. 2, 12489 Berlin, Germany.

[b] University of Bremen, Solid State Chemical Crystallography, Institute of Inorganic Chemistry and Crystallography, Leobener Str. 7, 28359 Bremen, Germany

[c] University of Bremen, MAPEX Center for Materials and Processes, Bibliothekstraße 1, 28359 Bremen, Germany

[d] Department of Materials Science and Engineering, Cornell University, Ithaca, NY 14853-1501, USA

[e] Platform for the Accelerated Realization, Analysis, & Discovery of Interface Materials (PARADIM), Cornell University, Ithaca, NY 14853-1501, USA

[f] Quest Integrated, LLC, 19823 58th Place S, Suite 200, Kent, WA 98032-2183, USA

[g] Kavli Institute at Cornell for Nanoscale Science, Ithaca, NY 14853, USA

Abstract

Large single crystals of the double-perovskite $Ba_2ScNbO_6$ were grown from the melt for the first time. With a lattice parameter at room temperature of 4.11672(1) Å, this cubic double-perovskite has an excellent lattice match to $BaSnO_3$, $PbZr_{0.9}Ti_{0.1}O_3$, $LaInO_3$, $BiScO_3$, and other perovskites of contemporary interest. Differential thermal analysis showed that $Ba_2ScNbO_6$ melts at 2165±30 °C in an inert atmosphere. Competitive grain growth was visualized by energy dispersive Laue mapping. X-ray diffraction rocking curve measurements revealed full width at half maximum values between 21 and 33 arcsec for 002 and 004 reflections. The crystals were sufficiently large to yield (100)-oriented single-crystal substrates with surface areas as large as 10 x 10 $mm^2$.

* Corresponding author. E-mail address: christo.guguschev@ikz-berlin.de (C. Guguschev)

1. Introduction

In recent years BaSnO$_3$ has emerged as a transparent conducting oxide (TCO) with excellent properties. Its high mobility at room temperature (over 300 cm$^2$V$^{-1}$s$^{-1}$ when degenerately doped [1]), excellent transparency, chemical stability, and indium-free composition make BaSnO$_3$ of great interest for future oxide electronic devices [2]. Already all-oxide field-effect transistors with BaSnO$_3$ channels have been demonstrated that operate at room temperature with peak field-effect mobilities of 61 cm$^2$V$^{-1}$s$^{-1}$ and I$_{on}$/I$_{off}$ ratios up to 10$^9$ [3].

A key limitation to electronic devices based on BaSnO$_3$ thin films, however, is the lack of a suitable substrate. Although epitaxial thin films of BaSnO$_3$ have been grown on a variety of substrates, so far they all suffer from high concentrations of threading dislocations with a typical density near 10$^{11}$ cm$^{-2}$ for BaSnO$_3$ films that are 100-300 nm thick [4, 5]. Films of this thickness are used in today's proof-of-principle BaSnO$_3$-based devices. This high density of dislocations is considered responsible for the significantly lower mobilities seen in thin films of BaSnO$_3$ compared with single crystals [1, 4, 6] the highest mobility reported to date in a BaSnO$_3$ thin film is 183 cm$^2$V$^{-1}$s$^{-1}$ at room temperature [5]. Although there are now nearly 20 commercially available single crystal substrates that are isostructural with BaSnO$_3$ (all perovskites [7]), none of them are well lattice matched to BaSnO$_3$. The closest is PrScO$_3$, which is mismatched by –2.3% [8]. When BaSnO$_3$ films are grown on these substrates, the inescapable result for films with device-relevant thicknesses is that the films are riddled with threading dislocations.

Perovskite single crystals with a better lattice match to BaSnO$_3$ are being investigated as potential substrates. These include LaInO$_3$ [9], (LaLuO$_3$)$_{1-x}$(LaScO$_3$)$_x$ solid solutions [10], and additional suggestions including SrZrO$_3$ [11]. The growth of large BaSnO$_3$ single crystals, which would enable homoepitaxial BaSnO$_3$ films, is also being pursued [12, 13], but to date the largest substrates from such crystals are 5 mm x 5 mm in size [13].

Identifying a suitable substrate for the growth of BaSnO$_3$ is a formidable challenge. The ideal substrate must satisfy many criteria including: (1) chemical compatibility, (2) structural compatibility, and (3) the ability to be grown as a large single crystal with excellent structural perfection [14, 15]. This final criterion is the most restrictive as there are relatively few crystal growth techniques capable of growing large single crystals with excellent structural perfection. The gold standard of these techniques is the Czochralski method, which in addition to its importance for conventional semiconductors is also the leading technique for large-area perovskite oxide substrates with the highest structural quality [16, 17]. Oxide crystals that show very intense infrared absorption and very low thermal conductivity at high temperatures suffer from growth instabilities like diameter fluctuation, foot formation and subsequent spiraling shortly after the seeding stage, which make automated process control during crystal growth very challenging to utilize. For such materials, alternative techniques like the edge-defined film-fed growth (EFG) method, directional solidification or the top-seeded solution growth (TSSG) method can be important backups to achieve decent crystal volumes suitable for small-scale fabrication of novel high-quality substrate crystals.

Recognizing the challenge in finding an $ABO_3$ perovskite (a simple perovskite) that can meet the above conditions and thus serve as a high quality substrate for $BaSnO_3$, we broadened our search criteria to include double-perovskite oxides with general formula $A_2BB'O_6$. Only 278 oxides with the simple $ABO_3$ perovskite structure are known [18] whereas 899 oxides with the double-perovskite structure are known [18] and far more (over 62,000) double-perovskite oxides have been calculated to satisfy the Goldschmidt criteria [19] and thus have the possibility of forming the perovskite structure (though many of these latter potential double-perovskites are metastable) [18]. Here we investigate the double-perovskite $Ba_2ScNbO_6$ as a potential substrate for $BaSnO_3$. This choice is motivated by the excellent structural compatibility between $Ba_2ScNbO_6$ and $BaSnO_3$: both have the perovskite structure with identical Ba-O layers and are lattice matched within better than 0.1%. Further, $Ba_2ScNbO_6$ has been reported to melt congruently at about 2130 °C [20], to be cubic [21-24], and to have a dielectric constant of 16 (Ref. [25]) and a bandgap of 3.6 eV [26]. Thus, $Ba_2ScNbO_6$ satisfies many important constraints to be a good substrate.

In this communication we describe the growth of the first large single crystals of $Ba_2ScNbO_6$, including the challenges that we encountered and overcame.

2. Experimental

2.1 Bulk single crystal growth

For the preparation of the starting materials, dried powders of $BaCO_3$, $Nb_2O_5$, $Sc_2O_3$ and MgO with purities of 99.99% (4N) were used. MgO was added with the intention to reduce the free carrier absorption at high temperatures by a mechanism known as compensation doping. The powders were weighed, mixed and calcined (in air at a 1300 °C for 12 h) to produce the following nominal starting composition: 66.45% BaO, 16.28% $Sc_2O_3$, 16.61% $Nb_2O_5$, 0.66% MgO (all given values are in mol%). Note that this nominal composition corresponds to doping the perovskite $B$-site with 1 at% Mg. Subsequently, cylindrical bars of this powder were made by cold isostatic pressing at 0.2 GPa to optimize the crucible filling process. The crystal growth experiments were performed using a conventional RF-heated Czochralski set-up. The atmosphere during growth was argon under ambient pressure. Iridium crucibles (inner diameter: 38 mm) embedded in $ZrO_2$ and $Al_2O_3$ insulation were used. Additional thermal insulation was added to the central region just above the surface of the melt. This was done (1) to allow nucleation of $Ba_2ScNbO_6$ crystals at the crucible wall and (2) to maintain grain-selection and continuous grain enlargement towards the central part of the crucible during cooling of the growth setup. In the work described, extensive Czochralski growth experiments were not performed due to the appearance of very intense growth instabilities.

### 2.2  EDLM and XRD measurements

Crystal quality was evaluated using two complementary methods. Energy dispersive Laue mapping (EDLM) provided a qualitative and non-destructive visualization of the large single crystalline area achieved by the grain selection process. Additionally, X-ray diffraction (XRD) measurements were performed to verify the crystalline quality of polished (001) $Ba_2ScNbO_6$ substrates using a four-circle XRD employing $CuK\alpha_1$ radiation. This was performed with a high-resolution (PANalytical X'Pert Pro MRD with a PreFix hybrid 4×Ge 220 monochromator on the incident beam side and a triple axis/rocking curve attachment (Ge 220) on the diffracted beam side). $\theta$-$2\theta$ scans and rocking curves of the 002 $Ba_2ScNbO_6$ and the 004 $Ba_2ScNbO_6$ substrate peaks were measured in a triple-axis geometry. EDLM mappings were performed at a pressure of about 1-2 mbar using a μ-XRF spectrometer M4 Tornado (Bruker) on the as-grown surface of the grown crystal. The measurement system was equipped with a rhodium X-ray source operated at 50 kV and 600 μA. Polycapillary X-ray optics were used to focus the non-polarized white radiation at the surface of the sample, resulting in a spatial resolution of about 20 μm. The white radiation from the excitation source interacts with the crystals and due to the instrument geometry it is possible to detect Bragg reflections. Using a measurement time of 10 ms per point, the sample surface was scanned "on the fly" by moving the sample stage in 30 μm step increments to create two-dimensional diffraction intensity maps of selected Bragg peaks. To increase count statistics, two passes of the scans were performed. The principle of the method, the measurement procedure and the measurement setup are described in detail elsewhere [27].

X-ray powder diffraction data of a ground up piece of a single crystal were collected on an X'Pert MPD PRO diffractometer (PANalytical GmbH, Almelo, The Netherlands) equipped with Ni-filtered $CuK_{\alpha 1,2}$ radiation ($\lambda_{K\alpha 1}$ = 154.05929(5) pm, $\lambda_{K\alpha 2}$ = 154.4414(2) pm) and a X'Celerator detector system in Bragg-Brentano geometry. Room-temperature scans were performed from 5 to 130° $2\theta$ with a step width of 0.0167° $2\theta$ and a measurement time of 140 s per step. The Rietveld refinements were carried out using "DiffracPlus Topas 4.2" (Bruker AXS GmbH, Karlsruhe, Germany) software combined with a fundamental parameter approach. The fundamental parameter set was determined by fitting the instrumental parameters against a $LaB_6$ standard reference material, which was verified by a silicon standard reference sample. In this way, precise lattice parameters could be determined that were used also for the final refinements of the single crystal data.

The single crystal diffraction (SCXRD) was performed on a D8 Venture (Bruker AXS GmbH, Karlsruhe, Germany) using Mo$_{K\alpha}$ radiation ($\lambda_{K\alpha 1}$ = 70.93171(4) pm, $\lambda_{K\alpha 2}$ = 71.3607(12) pm). The instrument was equipped with a KAPPA four–circle goniometer, a PHOTON 100 CMOS detector (active area 100 cm$^2$) and a curved graphite crystal TRIUMPH monochromator. A part of the original large crystal (with an ICP-OES determined magnesium content of about 100 wt ppm) was crushed into small pieces and a fragment of 71 x 98 x 158 µm$^3$ was fixed on a glass fiber mounted on a goniometer head. A full sphere of data up to a maximum resolution of 50 pm was collected at room temperature with a frame width of 0.5°. The frames were integrated using the SAINT software (Bruker AXS GmbH, Karlsruhe, Germany) and the intensities corrected for absorption using a numerical approach. Space group determination was carried out by analyzing the systematically absent reflections using the Xprep [28] software. The structure solution and refinement were performed using the SHELX software suite [28]. In the final steps of refinement, anisotropic displacement parameters for all sites, extinction and the suggested weighing scheme were applied, including the lattice parameters obtained from powder diffraction data Rietveld refinements.

### 2.3 Thermal analysis methods

The melting behavior of Ba$_2$ScNbO$_6$ was investigated by differential thermal analysis with simultaneous thermogravimetry (DTA/TG) using a NETZSCH STA 429CD. The sample was placed in a tungsten crucible with a lid. Prior to heating, the chamber was evacuated three times to a pressure of a few 10$^{-6}$ mbar and refilled with helium (99.9999% purity) between evacuations. The measurements were performed under static atmosphere after refilling the chamber with helium up to the normal pressure of 1 atm. Two subsequent heating/cooling cycles with rates of 15 K/min were performed.

The lack of good substances for temperature calibration for DTA measurements at such high temperatures is a well known problem. This is because reliable melting points are only known for a few metals from the literature, and molten metals cannot be simultaneously present in the tungsten crucibles during the DTA measurements of Ba$_2$ScNbO$_6$. For this investigation, a sample of pure Al$_2$O$_3$ (melting point 2054°C) [29] was molten under the same conditions prior to the Ba$_2$ScNbO$_6$ sample and the observed melting peak was used for a rough calibration of temperature. Nevertheless, the temperature error is expected to be of order ±30 K.

3. Results and discussion

3.1 Crystal growth

Ba$_2$ScNbO$_6$ single crystals were grown from the melt using iridium crucibles. The largest single crystal we achieved had dimensions of about 17 mm x 17 mm x 15 mm. This single-crystal is shown in Fig. 1 as part of a larger multicrystalline volume. The single crystal area is visualized by the color-coded intensity plot (Fig. 2a) of a selected Bragg reflection (Fig. 2c). Except for minor intensity variations caused by topographic effects, no subgrains or grains were detected by using EDLM on the relatively flat as-grown surface (Fig. 2b). Reconstruction of how the crystal might have grown suggests that the large grain expanded towards the central upper part of the crucible after nucleating near or at the crucible wall (Fig. 1a-b). Most of the crystal volume was colorless and water-clear; only the rim part of the crystal appears slightly bluish. From the large single-crystal part wafers of various sizes were prepared by CrysTec GmbH (Berlin, Germany). Figure 3 shows a differential interference contrast (DIC) micrograph of a representative wafer with a surface area of 10 x 10 mm².

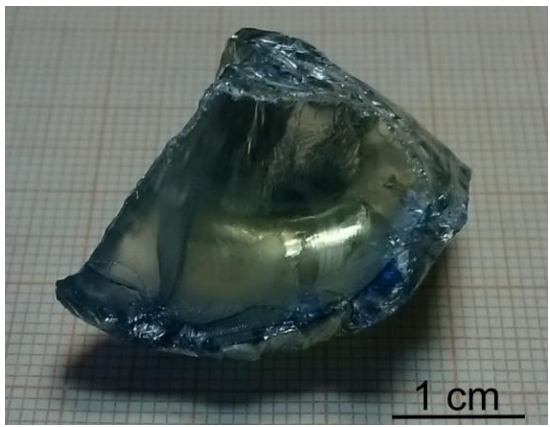

Fig. 1  Ba$_2$ScNbO$_6$ single crystal containing multicrystalline regions at the rim.

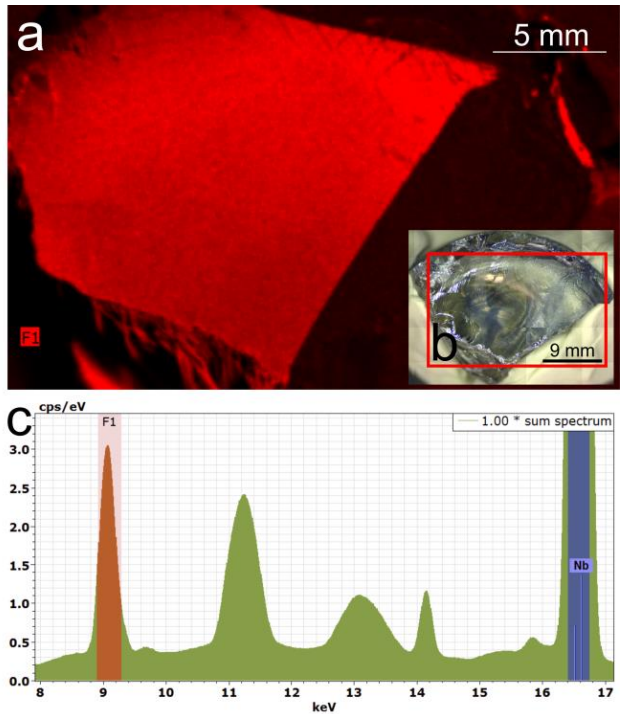

Fig. 2 (a) The color-coded intensity plot of a Bragg reflection (F1) represents the single-crystal region (colored red) of the $Ba_2ScNbO_6$ crystal. Slight differences in Bragg peak intensity are related to topographic effects, since the crystal was measured in the as-grown state. The measurements were performed within the area indicated by the red box (b) in the mosaic image and the sum spectra collected within the single-crystal area are shown in (c).

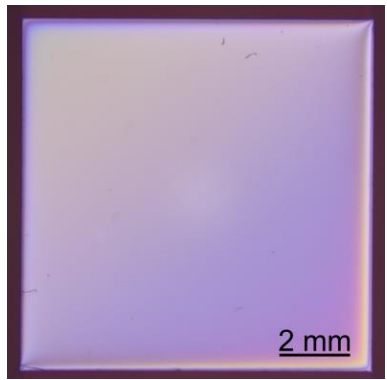

Fig. 3 Differential interference contrast micrograph of a chemo-mechanically polished (100)-oriented single-crystal wafer of $Ba_2ScNbO_6$ with a surface area of 10 x 10 $mm^2$ prepared from the crystal shown in Fig. 1. The dark lines are caused by dust particles sitting on the surface. The substrates were oriented, cut and polished by CrysTec, Berlin.

## 3.2   Structural quality and surface roughness

Investigation of the structural quality by XRD diffraction measurements confirmed the high crystalline quality within the single-crystal region described above. A $\theta$-$2\theta$ XRD scan of a polished (001)-oriented $Ba_2ScNbO_6$ substrate is shown in Fig. 4. Only the expected 00$\ell$ peaks are seen and these peaks are extremely sharp. No peaks indicating ordering of the unit cell beyond the ~4 Å unit cell of a simple $ABO_3$ perovskite are seen, despite this being a double-perovskite. This is in contrast to studies of bulk $Ba_2ScNbO_6$ powder where partial ordering has been observed corresponding to a doubling of the cubic $a$-axis lattice parameter [21, 22, 24]. Perhaps such ordering is favored at temperatures well below the melting temperature, i.e., temperatures through which the crystals in the present study traversed relatively quickly.

Rocking curves (in $\omega$) of the 002 and 004 peaks of a polished (001) $Ba_2ScNbO_6$ substrate are shown in Fig. 5. With FWHM in the 21-33 arc sec range, the $Ba_2ScNbO_6$ substrates are seen to be of high structural quality.

An AFM image of the surface of a polished (001) $Ba_2ScNbO_6$ substrate is shown in Fig. 6. The surface is smooth, but atomic steps are not seen. This is expected since this is a chemo-mechanically polished surface. To see steps a surface termination etch or anneal (or both) will need to be developed as have been developed for other perovskite substrates [30-33].

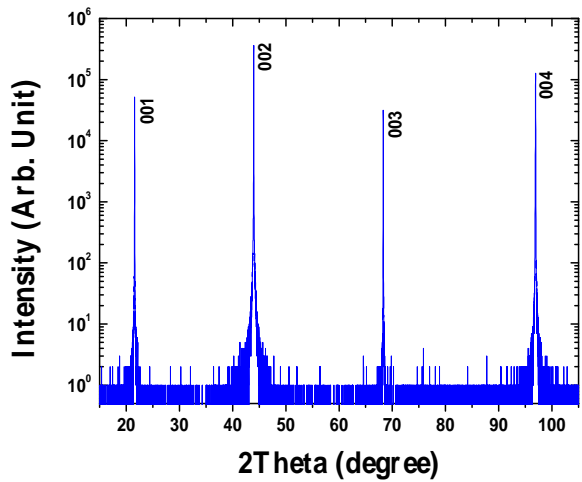

Fig. 4   $\theta$-$2\theta$ XRD scan of a polished (001)-oriented $Ba_2ScNbO_6$ substrate.

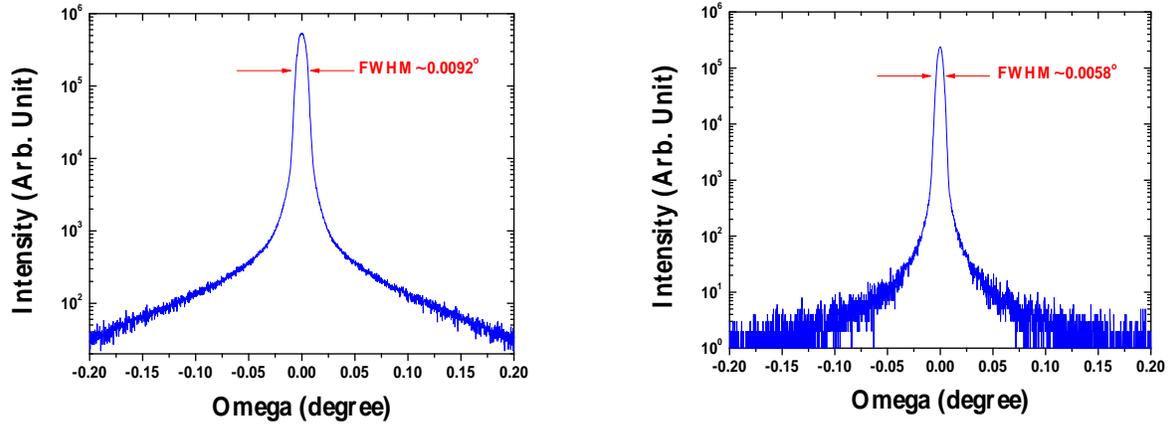

Fig. 5   Measured rocking curves (in $\omega$) of the 002 (at the left) and 004 (at the right) peak of a polished (001) $Ba_2ScNbO_6$ substrate.  The FWHM of these peaks vary from 21 to 33 arc sec.

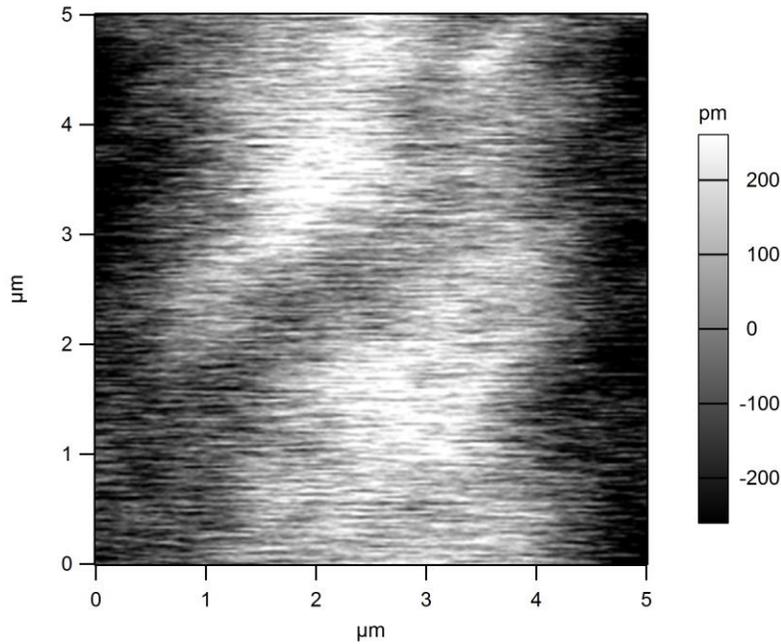

Fig. 6   AFM image of the surface of a polished (001) $Ba_2ScNbO_6$ substrate.

### 3.3 Lattice parameter / structural data

Evaluation and refinement of the single crystal diffraction data confirmed the cubic perovskite type $ABO_3$ structure. Barium was found on the *A* position whereas niobium and scandium both randomly occupied the *B* position in the center of the unit cell. Refinement of the occupancy parameters showed a slightly increased scandium content on this position, which could not be confirmed by the powder data Rietveld refinements where an equal distribution was found within 2 standard deviations. Nevertheless, also here a minor increase in the scandium content could be refined. The powder data were collected to obtain more accurate lattice parameters than possible from single crystal data (without measuring to very low *d*-values). These lattice parameters were calculated using Rietveld refinements using a fundamental parameter approach where the fundamental parameters are calibrated against an $LaB_6$ standard, which was then refined against a silicon standard, to ensure the highest accuracy of the resulting data. The crystallographic data and structure refinement obtained are given in Tables 1 and 2, respectively. The Rietveld plot is given in Fig. 7. For the final refinement of the structure from the single crystal data, the lattice parameters obtained from room-temperature powder data refinements were used due to their higher accuracy. Doing so the refined interatomic Ba-O distances of 291.36(2) pm are within the expected range as well as are the Sc/Nb-O distances of 206.02(1) pm, which must correspond to the Ba-O interatomic distance divided by the square root of 2 due to the symmetry relations of the cubic perovskite structure.

Table 1: Single crystal and powder diffraction data refinement results and crystallographic data of $Ba_2ScNbO_6$

| | Single-crystal data | Powder data |
|---|---|---|
| Symmetry | Cubic | |
| Space group | $Pm\bar{3}m$ (221) | |
| Lattice parameter a /pm | 412.04(2)* | 411.672(1) |
| Formula Weight /g mol$^{-1}$ | 254.27(1) | |
| Density /g cm$^{-3}$ | 6.036(1) | 6.04464(6) |
| h, k, l / data range $2\theta$/° | -8 < h,k,l < 8 | 5 - 130 |
| Reflections/data collected | 4647 | 7480 |
| Unique reflections/ data used | 146 | 6882 |
| $R_{int}$/$R_{exp}$ | 0.0136 | 0.0338 |
| $R_1$ (all) / $R_B$ | 0.0109 | 0.0451 |
| $wR_2$ / $R_{WP}$ | 0.0258 | 0.0790 |
| $R_P$ | | 0.0575 |
| GooF | 1.316 | 2.34 |
| Rest electron density /e$^-$ m$^{-30}$ | 0.71/-1.16 | |
| Ba-O /pm | 291.36(2) | |
| Nb/Sc-O /pm | 206.02(1) | |

* powder data lattice parameter used for structure refinements

Table 2: Atomic parameters of $Ba_2ScNbO_6$. Values obtained from powder data are given in italics.

| Atom | Wyckoff | x | y | z | Occup. | $U_{11}$[c] /pm$^2$ | $U_{22} = U_{33}$ /pm$^2$ | $U_{eq}$ /pm$^2$; $U_{iso}$ /pm$^2$ |
|---|---|---|---|---|---|---|---|---|
| Ba | 1a | 0 | 0 | 0 | 1 | 86(1) | = $U_{11}$ | 86(1) *16(2)* |
| Sc | 1b | ½ | ½ | ½ | 0.541(8) *0.506(3)* | 60(2)[a] | = $U_{11}$ | 60(2) *20(3)[b]* |
| Nb | 1b | ½ | ½ | ½ | 0.439(8) *0.494(3)* | 60(2)[a] | = $U_{11}$ | 60(2) *20(3)[b]* |
| O | 3c | 0 | ½ | ½ | 1 | 105(5) | 129(4) | 121(3) *99(4)* |

[a,b]values have been constrained refined; [c] $U_{12} = U_{23} = U_{13} = 0$

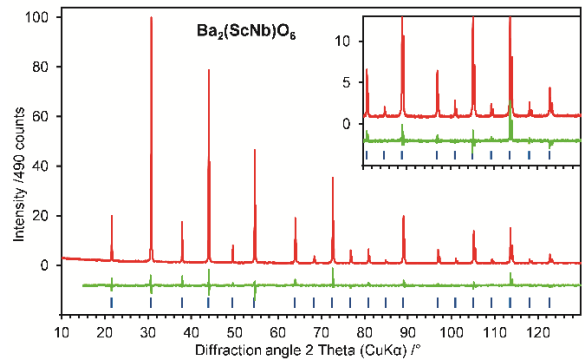

Figure 7: Rietveld plot of $Ba_2ScNbO_6$.

3.4 Melting behavior

An optically clear piece of Mg-doped $Ba_2ScNbO_6$ (mass ca. 100 mg) was heated ca. 15 K above the end of the melting peak that is shown in Fig. 8. It is obvious that the DTA basis, which is indicated as a dotted line, turns strongly in the endothermal direction, starting at the melting peak onset. It can be assumed that the endothermal effect results from the initiation of the evaporation of material, which amounts to ca. 1 % by weight in this first heating and which consumes heat.
At the high-temperature extreme of this DTA/TG measurement, which extends 2200 °C, the binary oxide components of $Ba_2ScNbO_6$ with the highest vapor pressures are BaO ($p$=20 mbar), Mg ($p$=1 mbar, evaporates as metal) and $NbO_2$ ($p$=0.5 mbar) [29], whereas the vapor pressure of all scandium species is less than $10^{-6}$ bar. Consequently, depletion of the sample in Ba and oxygen (and to a smaller degree possibly in Nb and the dopant Mg) must be expected during melting. During cooling a strong exothermal peak (not shown in Fig. 7) indicates crystallization near 2095 °C, which means ca. 70 K supercooling below the melting point. In the second heating run melting starts almost at that same lower temperature, probably a result of the previous concentration shift, which affects subsequent melting events significantly.

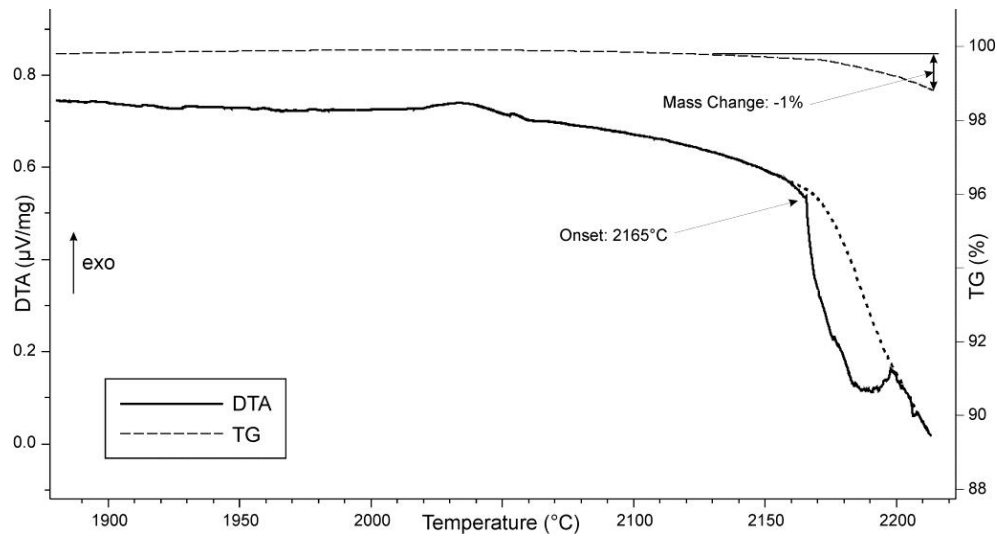

Fig. 8: The melting of $Ba_2ScNbO_6$ at $(2165\pm30)°C$ is accompanied by a significant mass loss of 1%

4. Conclusions and outlook

Large high-quality $Ba_2ScNbO_6$ single crystals with dimensions of up to 17 mm x 17 mm x 15 mm were grown from the melt for the first time using an unconventional and in some sense novel crystal growth approach. This novel substrate crystal with cubic perovskite type structure and a lattice parameter of 4.11672(1) Å offers excellent lattice match to several perovskites of contemporary interest. Due to its potential for a series of applications, a patent application is ongoing (US reference number: 16/424,987), that also includes details of the crystal growth technique. In order to produce larger $Ba_2ScNbO_6$ single crystals future work will be focused on finding suitable growth windows for seeded growth runs. To undertake this kind development, detailed knowledge of the phase relations in the relevant part of $BaO-Sc_2O_3-Nb_2O_5$ system is crucial and therefore currently under investigation.

Acknowledgements

The authors acknowledge fruitful discussions with Pat Woodward regarding the properties of potential double perovskite substrates. The work at Cornell was supported by the Air Force Office of Scientific Research under Award No. FA9550-16-1-0192 and by the National Science Foundation [Platform for the Accelerated Realization, Analysis, and Discovery of Interface Materials (PARADIM)] under Cooperative Agreement No. DMR-1539918. Substrate preparation was performed in part at the Cornell NanoScale Facility, a member of the National Nanotechnology Coordinated Infrastructure (NNCI), which is supported by the NSF (Grant No. ECCS-1542081).